\documentclass[english]{iopart}
\usepackage[T1]{fontenc}
\usepackage[latin9]{inputenc}
\usepackage{array}
\usepackage{graphicx}

\makeatletter

\providecommand{\tabularnewline}{\\}

\usepackage{iopams}
\usepackage{setstack}
\usepackage{bm,pstricks,psfrag}

\usepackage{babel}
\makeatother

\begin{document}

\title{An Euler Poincaré framework for the multilayer Green Nagdhi equations}

\author{J. R. Percival$^{1}$, C. J. Cotter$^{2}$ and D. D. Holm$^{1}$}

\address{$^{1}$Mathematics Department, Imperial College London, SW7 2AZ,
UK }

\address{$^{2}$Aeronautics Department, Imperial College London, SW7 2AZ,
UK}

\ead{j.percival@imperial.ac.uk}

\begin{abstract}
The Green Nagdhi equations are frequently used as a model of the wave-like
behaviour of the free surface of a fluid, or the interface between
two homogeneous fluids of differing densities. Here we show that their
multilayer extension arises naturally from a framework based on the
Euler Poincaré theory under an ansatz of columnar motion. The framework
also extends to the travelling wave solutions of the equations. We
present numerical solutions of the travelling wave problem in a number
of flow regimes. We find that the free surface and multilayer waves
can exhibit intriguing differences compared to the results of single
layer or rigid lid models.
\end{abstract}
\maketitle

\section{Introduction}

Internal gravity waves have been observed propagating in many different
locations in the world's oceans, both through direct measurement of
the change in density stratification \cite{Ramp2004} and through
the observed change in surface conditions, from satellite observations
\cite{ZKZY2003}and from recent observations from the Space Shuttle
in the region of Dongsha in the South China Sea \cite{LCHL98}. These
waves play a key role in the transport of energy in the ocean and,
through wave breaking, help to control mixing \cite{EMPS2006}. The
observed waves are not classical one dimensional phenomena. With wave
crests extending up to 200 km perpendicular to the direction of motion,
the wave properties vary with depth, exhibit curvature and interact
with the underlying bathymetry and with each other. These interactions
give rise to a number of phenomena including refraction, diffraction
and wave front reconnection.

This paper examines one strongly nonlinear, multilayer, two dimensional
equation set for the behaviour of such waves, which is derivable from
an Euler Poincaré variational principle and shows that the equations
admit travelling wave solutions which exhibit many interesting phenomena.
We begin by discussing the derivation of the multilayer Green Nagdhi
equations from a variational principle under the constraint of columnar
motion. We then show the dynamics contain a strong, fast barotropic
mode and relate this to the rigid lid models of other investigators.
Finally we present numerical solutions to the travelling wave problem
for the system, illustrating the important behaviour of the free surface.

\section{The multilayer Green Nagdhi equations}

\begin{figure}[bh]
\begin{centering}
\includegraphics[width=0.6\columnwidth]{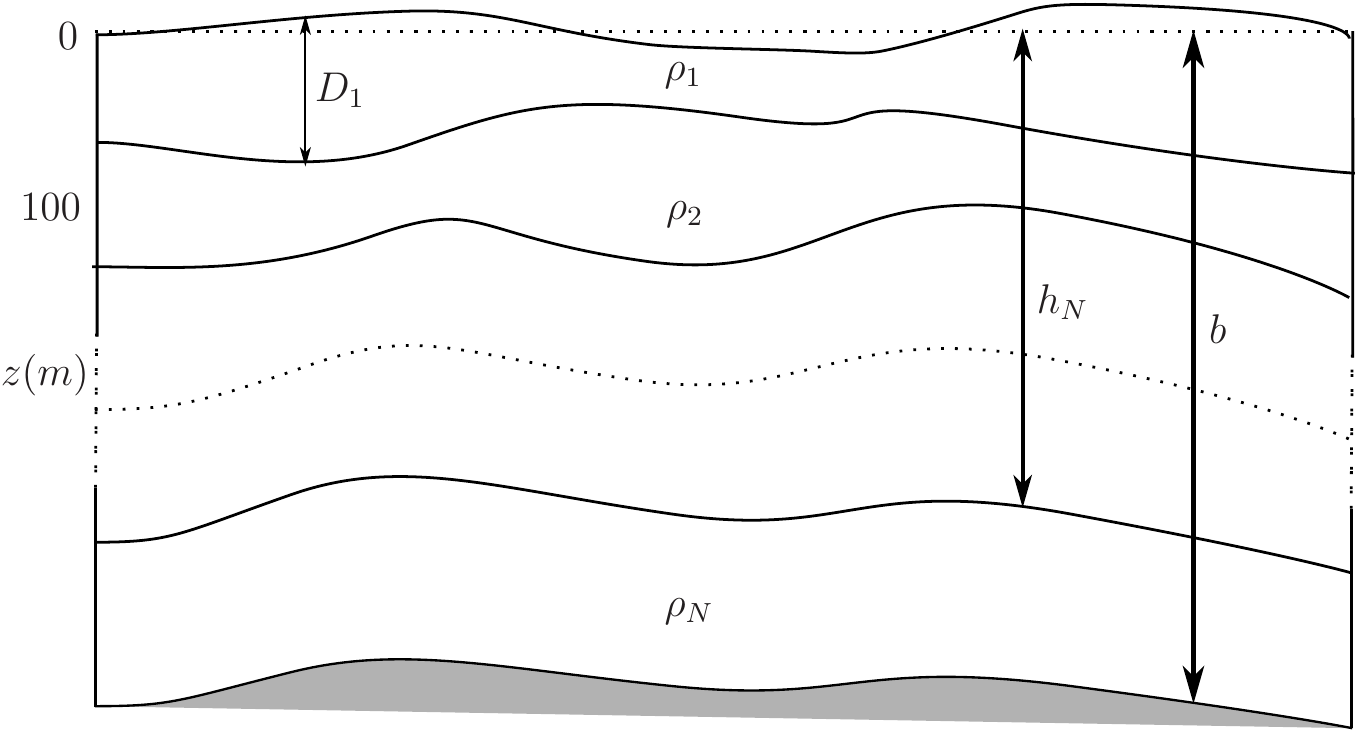}\caption{The multilayer fluid system}
\label{fig:MLS}
\par\end{centering}
\end{figure}

We will begin by deriving a multilayer extension of the shallow water
wave equations commonly attributed to Green and Nagdhi \cite{GN74},
using a variational principle under the ansatz of columnar motion.
Consider a system of $N$ homogeneous fluid layers which are acted
upon by a constant gravitational acceleration, $g$ and with the layers
initially arranged in a stable stratification. Each layer thus possesses
a constant density, $\rho_{i}$ for $i=1,\ldots,N$, with $\rho_{1}<\rho_{2}<\ldots<\rho_{N}$.
We have chosen by convention to number the layers downwards from the
free surface (see Figure \ref{fig:MLS}). Let $(\mathbf{u}_{i},w_{i})$
denote the full three dimensional velocity fields within the layers
and $h_{i}$ denote the depth of the layer interfaces with respect
to the mean free-surface height at $z=0.$ We define $b=-h_{N+1}$
to be the fixed bottom topography to which the lowest layer is assumed
to remain attached. Summing over the layers, the total kinetic energy
of the system is

\[
\int Kd\mathbf{x}:=\sum_{i=1}^{N}\int\int\frac{\rho_{i}}{2}\int_{h_{i+1}}^{h_{i}}|\mathbf{u}_{i}|^{2}+w_{i}^{2}dzd\mathbf{x},\]
while the total gravitational potential energy, relative to a background
state with vanishing density, is

\[
\int Vd\mathbf{x}:=\int\sum_{i=1}^{N}\rho_{i}\int_{h_{i+1}}^{h_{i}}zdz\mathbf{dx}=\int\sum_{i=1}^{N}\frac{\rho_{i}}{2}(h_{i}^{2}-h_{i+1}^{2})d\mathbf{x}.\]
We also assume two auxiliary equations within each layer, namely a
three dimensional incompressibility condition,\begin{equation}
\nabla\cdot\mathbf{u}_{i}+\frac{\partial w_{i}}{\partial z}=0,\label{eq:incomp}\end{equation}
and a transport equation for the layer thicknesses, $D_{i}=h_{i}-h_{i+1}$,
based upon the conservation of fluid within each layer,\begin{equation}
\frac{\partial D_{i}}{\partial t}+\nabla\cdot D_{i}\mathbf{u}_{i}=0.\label{eq:trnsp:D}\end{equation}

We now make the ansatz of columnar motion with respect to the horizontal
components of velocity in each of the layers, so that \begin{equation}
\frac{\partial\mathbf{u}_{i}}{\partial z}=0\quad{\rm for}\ i=1,\ldots,N.\label{eq:anzatz}\end{equation}
Together (\ref{eq:incomp}) and (\ref{eq:anzatz}) enforce a linear
dependence on $z$ for the vertical component of velocity, $w_{i}$,
so that using the bottom boundary condition,\[
\mathbf{u}_{N}\cdot\nabla\mathbf{b}+w_{N}=0,\]
 and integrating upwards we can find the vertical velocity within
each layer in terms of the horizontal velocity divergences within
that layer and those below it,

\begin{equation}
w_{i}=-z\nabla\cdot\mathbf{u}_{i}-\nabla\cdot b\mathbf{u}_{i}-\sum_{j=i+1}^{N}\nabla\cdot D_{j}(\mathbf{u}_{j}-\mathbf{u}_{i}).\label{eq:wdef}\end{equation}
This gives a velocity jump across the interfaces of\[
w_{i}(h_{i+1})-w_{i+1}(h_{i+1})=(\mathbf{u}_{i+1}-\mathbf{u}_{i})\cdot\nabla h_{i+1},\]
which is necessary to avoid the layers separating. Substituting (\ref{eq:wdef})
into the definition of kinetic energy and integrating in the vertical
gives

\[
\fl\int Kd\mathbf{x}=\int\sum_{i=1}^{N}\frac{\rho_{i}}{2}\left(D_{i}|\mathbf{u}_{i}|^{2}+D_{i}[\frac{D_{i}^{2}}{3}(\nabla\cdot\mathbf{u}_{i}){}^{2}+D_{i}\nabla\cdot\mathbf{u}_{i}W_{i}+W_{i}^{2}]\right)d\mathbf{x},\]
where we have defined quantities\[
W_{i}:=w_{i}(h_{i+1})=\mathbf{u}_{i}\cdot\nabla\left(-b+\sum_{j=i+1}^{N}D_{j}\right)+\sum_{j=i+1}^{N}\nabla\cdot D_{j}\mathbf{u}_{j},\]
which are equal to the vertical velocity at the bottom of each layer.
\[
\]

We now use Hamilton's principle in the form

\[
\delta S:=\delta\int\int\ell d\mathbf{x}dt=0,\]
applied for the reduced Lagrangian defined by

\[
\ell:=K-V,\]
coupled to (\ref{eq:trnsp:D}). The set of physically admissible horizontal
velocities in each layer (a collection of vector fields over the horizontal
domain, tangential on the boundary) form a Lie algebra under what
is termed the ideal fluid bracket operator for given for functions
$F,G$, of a vector field $\mathbf{v}_{i}$, by\begin{equation}
\{F,G\}(\mathbf{v}_{i}):=\int\mathbf{v}_{i}\cdot\left(\frac{\delta G}{\delta\mathbf{v}_{i}}\cdot\nabla\frac{\delta F}{\delta\mathbf{v}_{i}}-\frac{\delta F}{\delta\mathbf{v}_{i}}\cdot\nabla\frac{\delta G}{\delta\mathbf{v}_{i}}\right)d\mathbf{x},\label{eq:LieBracket}\end{equation}
for variations defined by \[
\lim_{\epsilon\rightarrow0}\frac{1}{\epsilon}\left[F(\mathbf{v}+\epsilon\delta\mathbf{v})-F(\mathbf{v})\right]=\int\delta\mathbf{v}\cdot\frac{\delta F}{\delta\mathbf{v}}d\mathbf{x},\]
while the layer thicknesses form a set of field densities which are
Lie-transported (i.e. advected) by the Eulerian velocity flow. This
places the system within the formalism of Euler-Poincaré theory on
semidirect products, which extends the results of Hamiltonian mechanics
into this more generalized algebraic structure. 

Similar formulations have previously been derived for several alternative
models for fluid flow, starting with the Euler equations, as originally
considered by Poincaré and including in particular the Camassa-Holm
\cite{Misiolek} and KdV \cite{Guha2002} shallow water equations,
models which, like the single layer Green Nagdhi equation, has been
used successfully to model the behaviour of nonlinearly dispersive
water waves. The Euler-Poincaré theory has connections with many other
topics in geometric mechanics, notably Lagrangian reduction \cite{CeMaRa2001},
which has been applied successfully in several areas of compressible
and incompressible fluid flow, and has implications for numerical
methods, through a more rigorous understanding of the flow of conserved
quantities in the system as geodesic motion on a suitable manifold.

Calculating the requisite adjoint actions and their duals to generate
the Euler-Poincaré equations \cite{HMR1998} in this framework (analogous
to the Euler-Lagrange equations for a finite dimensional Hamiltonian
system) gives the multilayer Green Nagdhi (MGN) equations of motion,

\begin{equation}
\frac{\partial\mathbf{m}_{i}}{\partial t}+\mathbf{u}_{i}\cdot\nabla\mathbf{m}_{i}+\mathbf{m}_{i}\cdot\nabla\mathbf{u}_{i}^{T}+\mathbf{m}_{i}\nabla\cdot\mathbf{u}_{i}=D_{i}\nabla\frac{\delta\ell}{\delta D_{i}},\label{eq:EP}\end{equation}
in terms of the layer momenta, $\mathbf{m}_{i}:=\frac{\delta\ell}{\delta\mathbf{u}_{i}},$
dual to the layer velocities. The MGN equation is forced by a pressure
like term, $\frac{\delta\ell}{\delta D_{i}}$, containing the effects
of gravity and nonhydrostatic terms, which appears out of viewing
the relation between kinetic and potential energy in the system as
a semi-direct product over the combined space of layer velocity and
thickness fields. Explicitly, these terms are given by

\[
\fl\mathbf{m}_{i}=\rho_{i}\left[D_{i}\mathbf{u}_{i}-\nabla\left(\frac{D_{i}^{3}}{3}\nabla\cdot\mathbf{u}_{i}+\frac{W_{i}}{2}\right)-D_{i}\left(\frac{D_{i}^{2}}{2}\nabla\cdot\mathbf{u}_{i}+W_{i}\right)\nabla h_{i+1}\right]\]

\[
-D_{i}\nabla\sum_{j=1}^{i-1}\rho_{j}D_{j}\left(\frac{D_{i}}{2}\nabla\cdot\mathbf{u}_{j}+W_{j}\right),\]

\[
\fl\frac{\delta\ell}{\delta D_{i}}=\frac{|\mathbf{u}_{i}|^{2}}{2}-\rho_{i}gh_{1}-\sum_{j=1}^{i-1}(\rho_{j}-\rho_{i})D_{j}\]
\[
\fl\qquad\quad+\sum_{j=1}^{i-1}\rho_{j}\left[\mathbf{u}_{i}\cdot\nabla\left(\frac{D_{j}}{2}\nabla\cdot\mathbf{u}_{j}+W_{j}\right)+\nabla\cdot\left(\frac{D_{j}}{2}\nabla\cdot\mathbf{u}_{j}+W_{j}\right)(\mathbf{u}_{j}-\mathbf{u}_{i})\right].\]

Although complex, many of the terms in the equation are in balance.
This is perhaps most clearly seen by rewriting (\ref{eq:EP}) in the
form

\[
\frac{\partial}{\partial t}\left(\frac{\mathbf{m}_{i}}{D_{i}}\right)+\mathbf{u}_{i}\cdot\nabla\left(\frac{\mathbf{m}_{i}}{D_{i}}\right)+\left(\frac{\mathbf{m}_{i}}{D_{i}}\right)\cdot\nabla\mathbf{u}_{i}^{T}=\nabla\frac{\delta\ell}{\delta D_{i}}.\]
Passing the material derivative part way through the definition of
the layer momenta gives \begin{eqnarray*}
\fl\frac{d_{i}\mathbf{u}_{i}}{dt}=\underbrace{-g\nabla\left(h_{1}+\sum_{j=1}^{i-1}\frac{\rho_{j}-\rho_{i}}{\rho_{i}}D_{j}\right)}_{P_{h}}+\frac{1}{D_{1}}\nabla D_{i}^{2}\frac{d_{i}}{dt}\left(\frac{D_{i}\nabla\cdot\mathbf{u}_{i}}{3}+\frac{W_{i}}{2}\right)\\
+\frac{d_{i}}{dt}\left(\frac{D_{i}\nabla\cdot\mathbf{u}_{i}}{2}+W_{i}\right)\nabla h_{i+1}+\nabla\sum_{j=1}^{i-1}\frac{\rho_{j}}{\rho_{i}}D_{j}\frac{d_{j}}{dt}\left(\frac{D_{j}\nabla\cdot\mathbf{u}_{j}}{2}+W_{j}\right) &  & ,\end{eqnarray*}
identical to the equation set given for the same fluid system in the
two layer case by Choi and Camassa \cite{CC96JFM} (henceforth the
CC equation) by an asymptotic expansion method and by Liska and Wendroff
\cite{LB1997} from directly substituting the definition of $w_{i}$
in the Euler equations. It also agrees with the one dimensional multilayer
equation of Choi \cite{Choi2000}. The quantity labelled $P_{h}$
is a hydrostatic pressure, representing forcing from both deviations
to the free surface height and from variations in the thicknesses
of the layers above. Calculating the full commutator $\left[d_{i}/dt,\mathcal{L}_{ij}\mathbf{u}_{j}\right]$
for the symmetric second order elliptic operator defined by $\mathbf{m}_{i}=\mathcal{L}(D,b)_{ij}\mathbf{u}_{j}$,
the equations may also be written entirely in terms of gradients of
the horizontal velocity and layer thicknesses,\begin{eqnarray}
\fl\mathcal{L}(D,b)\frac{d_{i}\mathbf{u}_{i}}{dt}=-\rho_{i}g\nabla\left(h_{1}+\sum_{j=1}^{i-1}\frac{\rho_{j}-\rho_{i}}{\rho_{i}}D_{j}\right)-\frac{\rho_{i}}{D_{i}}\nabla\left(D_{i}^{2}\left[\frac{R_{i}}{3}+\frac{S_{i}}{2}\right]\right)\label{eq:smoothedsw}\\
+\left(\frac{R_{i}}{2}+S_{i}\right)\nabla h_{i+1}+\nabla\sum_{j=1}^{i-1}\frac{\rho_{j}}{\rho_{i}}D_{j}\left(\frac{R_{j}}{2}+S_{j}\right),\end{eqnarray}
where

\[
R_{i}:=D_{i}\left[(\nabla\cdot\mathbf{u}_{i})^{2}+\tr(\nabla\mathbf{u}_{i}^{T}\cdot\nabla\mathbf{u})\right],\]

\[
S_{i}:=-\left[(\mathbf{u}_{i}\cdot\nabla\nabla h_{i+1})\cdot\mathbf{u}_{i}+\sum_{j=i+1}^{N}\nabla\cdot(\nabla\cdot(D_{j}\mathbf{u}_{j}\mathbf{u}_{j}))\right].\]
In this form we see the operator $\mathcal{L}$ acting as a smoother
on the hydrostatic pressure, while there is an induced remainder term
which grows with the non-linearity and further couples the layers.

A final rearrangement of (\ref{eq:EP}) gives an equation similar
to the vorticity form of the Euler equations,\[
\frac{\partial}{\partial t}\left(\frac{\mathbf{m}_{1}}{D_{1}}\right)+\nabla\left(\frac{\mathbf{u}_{i}\cdot\mathbf{m}_{i}}{D_{i}}\right)+\left[\nabla\times\left(\frac{\mathbf{m}_{i}}{D_{i}}\right)\right]\times\mathbf{u}_{i}=\nabla\frac{\delta\ell}{\delta D_{i}}.\]
Taking curls shows that the equations materially conserve a quantity,

\[
q_{i}:=\frac{1}{D_{i}}\nabla\times\left(\frac{\mathbf{m}_{i}}{D_{i}}\right),\]
identified with potential vorticity in each layer. These conservation
laws may also be derived from the existence of individual Kelvin circulation
theorems in each layer, namely \[
\frac{d}{dt}\oint_{c(\mathbf{u}_{i})}\frac{\mathbf{m}_{i}}{D_{i}}\cdot d\mathbf{x}=0,\]
where the integral is over the closed loop, $c,$ assumed to move
at the layer velocity, $\mathbf{u}_{i}$. In turn it can be shown
that the existance of these circulation theorems and their associated
conservation laws follows in turn from the invariance of the EP formulation
to fluid parcel relabelling in the configuration space, provided that
it preserves Eulerian quantities. For fuller details of the Kelvin-Noether
Theorem for EP equations in the context of semidirect product Lie
algebras see \cite{HMR1998}. The integral over the entire layer volume
of any function of the $q_{i}$ represents a conserved quantity of
the MGN equations. These are the Casmirs of the Lie-Poisson Hamiltonian
operator, and hence of the Lie-Poisson bracket, (\ref{eq:LieBracket}).
That is they are functionals, C, which satisfy $\{C,H\}=0$ for any
Hamiltonian, $H$.

\[
\]

\section{The barotropic and baroclinic modes}

\begin{figure}
\begin{centering}
\includegraphics[width=0.6\textwidth]{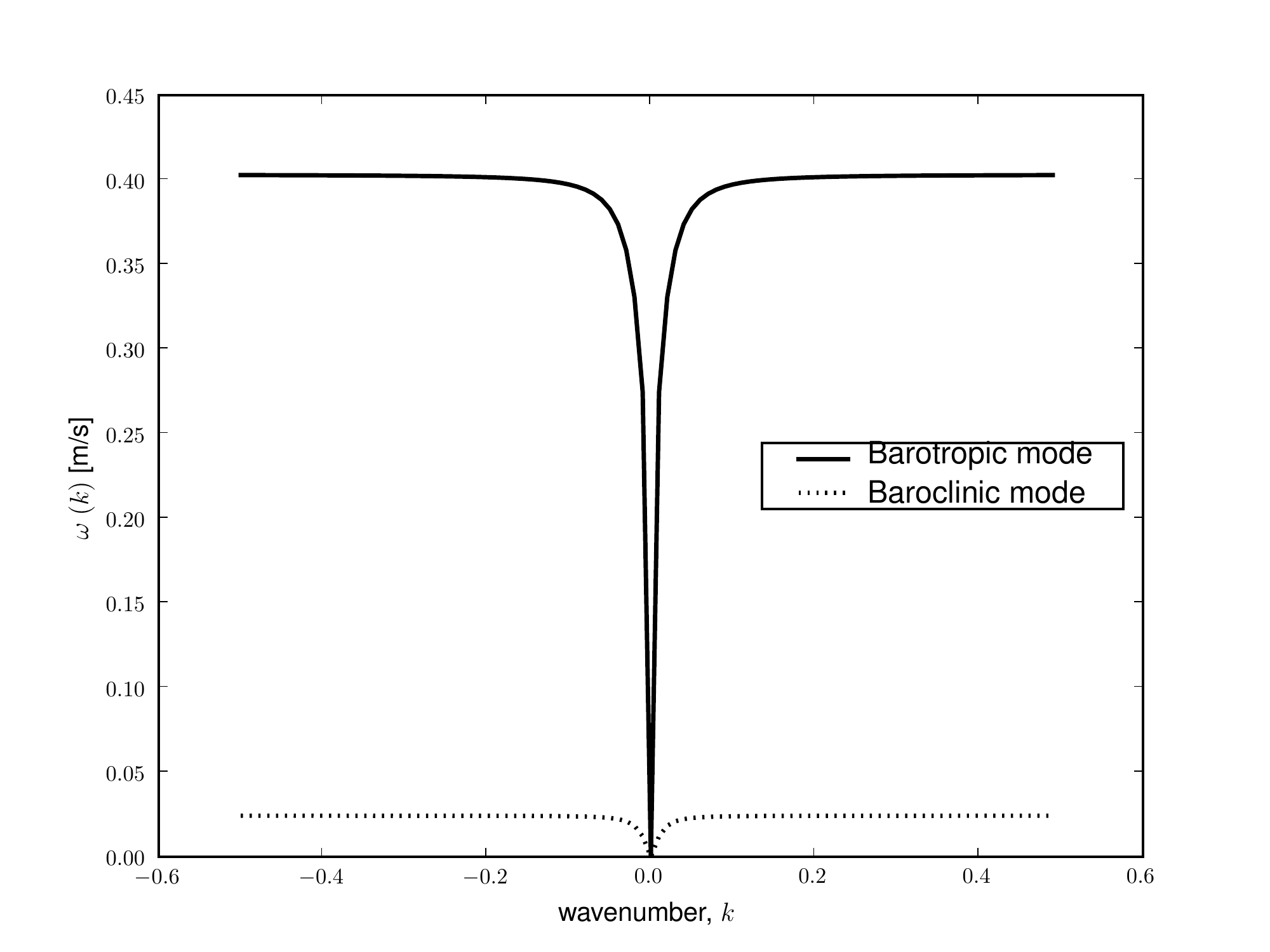}
\par\end{centering}

\caption{The linear dispersion relation for the MGN equations. In this case
$d_{1}=100$m, $\rho_{1}=1023$kg m$^{-3}$, $d_{2}=200$m, $\rho_{2}=1026$kg
m$^{-3}$. Note the wide separation of scales between the two modes. }
\label{fig:dispersion}
\end{figure}

Linearizing the 2-layer MGN equations around a state of rest, with
standing fluid heights $d_{1}$, $d_{2}$ gives the system 

\[
\frac{\partial}{\partial t}\left[\mathbf{u}_{1}-\frac{1}{3}d_{1}^{2}\nabla\nabla\cdot\mathbf{u}_{1}-\frac{1}{2}d_{1}d_{2}\nabla\nabla\cdot\mathbf{u}_{2}\right]=-g\nabla(D_{1}+D_{2}),\]

\begin{eqnarray*}
\fl\frac{\partial}{\partial t}\left[\mathbf{u}_{2}-\frac{1}{3}d_{2}^{2}\nabla\nabla\cdot\mathbf{u}_{2}\right.\\
-\left.\frac{\rho_{1}}{\rho_{2}}d_{1}\left(\frac{1}{2}d_{1}\nabla\nabla\cdot\mathbf{u}_{1}+d_{2}\nabla\nabla\cdot\mathbf{u}_{2}\right)\right]=-g\nabla\left(D_{2}+\frac{\rho_{1}}{\rho_{2}}D_{1}\right),\end{eqnarray*}

\[
\frac{\partial D_{1}}{\partial t}+d_{1}\nabla\cdot\mathbf{u}_{1}=0,\]

\[
\frac{\partial D_{2}}{\partial t}+d_{2}\nabla\cdot\mathbf{u}_{2}=0.\]
Taking divergences of the momentum evolution equations then give the
linearized wave equations

\[
\frac{\partial^{2}}{\partial t^{2}}\left(D_{1}-\frac{d_{1}^{2}}{3}\Delta D_{1}-\frac{d_{1}d_{2}}{2}\Delta D_{2}\right)=gd_{1}\Delta(D_{1}+D_{2}),\]
\begin{eqnarray*}
\fl\frac{\partial^{2}}{\partial t^{2}}\left[D_{2}-\frac{d_{2}^{2}}{3}\Delta D_{2}\right.\\
-\left.\frac{\rho_{1}d_{1}}{\rho_{2}}\Delta\left(\frac{d_{1}}{2}D_{1}+d_{2}D_{2}\right)\right]=gd_{2}\Delta\left(D_{2}+\frac{\rho_{1}}{\rho_{2}}D_{1}\right),\end{eqnarray*}
with $\Delta:=\partial^{2}/\partial x^{2}+\partial^{2}/\partial y^{2}$,
the standard two dimensional Laplacian operator. Specializing to one-dimensional
plane wave solutions, $D_{i}=c_{i}\exp[i(\omega t-kx)]$, we obtain
a linear dispersion relation\begin{eqnarray*}
\fl\left(1+\frac{1}{3}d_{1}^{2}k^{2}+\frac{1}{3}d_{2}^{2}k^{2}+\frac{\rho_{1}}{3\rho_{2}}d_{1}d_{2}k^{2}+\frac{1}{9}d_{1}^{2}d_{2}^{2}k^{4}+\frac{\rho_{1}}{12\rho_{2}}d_{1}^{3}d_{2}k^{4}\right)\omega^{4}\\
-gk^{2}\left((d_{1}+d_{2})\left[1+\frac{1}{3}d_{1}d_{2}k^{2}\right]-\frac{\rho_{1}}{2\rho_{2}}d_{1}(d_{1}-d_{2})^{2}k^{2}\right)\omega^{2}\\
+gg'd_{1}d_{2}k^{4}=0,\end{eqnarray*}
with $g':=g(\rho_{2}-\rho_{1})/\rho_{2}$ a reduced gravity term.
Plotting solutions shows both a fast barotropic mode, $\omega_{f}$,
and a slow baroclinic one, $\omega_{s}$, as shown in Figure \ref{fig:dispersion}.
Under the limits $g'd_{1}d_{2}\ll g(d_{1}+d_{2})^{2}$, $d_{i}k\ll1$
we re-obtain the shallow water modes,\[
\omega_{f}=\sqrt{g(d_{1}+d_{2})}k,\qquad\omega_{s}=\sqrt{g'\frac{d_{1}d_{2}}{d_{1}+d_{2}}}k.\]

Looking for one dimensional solutions to the linearized system with
$u_{2}=\lambda u_{1}$ and $D_{2}=(\mu-1)D_{1}$ we obtain equations

\begin{eqnarray*}
\frac{\partial}{\partial t}\left(1-\left[\frac{1}{3}d_{1}^{2}+\frac{\lambda}{2}d_{1}d_{2}\right]\frac{\partial^{2}}{\partial x^{2}}\right)u_{1}=-g\mu\frac{\partial D_{1}}{\partial x}\\
\frac{\partial}{\partial t}\left(\lambda-\left[\frac{\lambda}{3}d_{2}^{2}+\lambda\frac{\rho_{1}}{\rho_{2}}d_{1}d_{2}+\frac{1}{2}\frac{\rho_{1}}{\rho_{2}}d_{1}^{2}\right]\frac{\partial^{2}}{\partial x^{2}}\right)u_{1}=(-g\mu+g')\frac{\partial D_{1}}{\partial x}\\
\frac{\partial D_{1}}{\partial t}+d_{1}\frac{\partial u_{1}}{\partial x}=0\\
(\mu-1)\frac{\partial D_{1}}{\partial t}+\lambda d_{2}\frac{\partial u_{1}}{\partial x}=0\end{eqnarray*}

The final two equations give an obvious relation,

\[
(\mu-1)=\frac{d_{2}}{d_{1}}\lambda,\]
while the first two equations can be rearranged to give $\frac{\partial^{2}u}{\partial x^{2}}$
in terms of quantities other than $u$,

\begin{eqnarray*}
\fl\left[\frac{\lambda}{3}(d_{1}^{2}-d_{2}^{2})+\lambda\left(\frac{\lambda}{2}-\frac{\rho_{1}}{\rho_{2}}\right)d_{1}d_{2}\right.\\
-\left.\frac{1}{2}\frac{\rho_{1}}{\rho_{2}}d_{1}^{2}\right]\frac{\partial}{\partial t}\frac{\partial^{2}u_{1}}{\partial x^{2}}=(g\mu(\lambda-1)+g')\frac{\partial D_{1}}{\partial x}.\end{eqnarray*}
On re-substituting this gives a cubic equation for $\lambda$. For
$g'$ small one approximate solution is $\mu=1+d_{2}/d_{1},\lambda=1$,
another consistent solution has $\mu\approx0$, $\lambda\approx-d_{1}/d_{2}$.
We identify the first solution with the barotropic mode and the second
with the baroclinic. This shows the linear dynamics of the barotropic
and baroclinic modes are governed by

\[
\frac{\partial u_{bt}}{\partial t}=-g\frac{\partial D_{bt}}{\partial x},\qquad\frac{\partial D_{bt}}{\partial t}=(d_{1}+d_{2})\frac{\partial u_{bt}}{\partial x},\]
and

\[
\frac{\partial Lu_{bc}}{\partial t}=-g'\frac{\partial D_{bc}}{\partial x},\qquad\frac{\partial D_{bc}}{\partial t}+d_{1}\frac{\partial u_{bc}}{\partial x}=0,\]
respectively, where \[
Lu=\left(d_{1}/d_{2}-\left[d_{1}d_{2}/3+\frac{\rho_{1}}{\rho_{2}}d_{1}d_{2}-\frac{1}{2}\frac{\rho_{1}}{\rho_{2}}d_{1}^{2}\right]\frac{\partial^{2}}{\partial x^{2}}\right)u,\]
the linearization of the MGN operator. This shows that to leading
order the barotropic mode dynamics are precisely those of the one
layer shallow water equations. Deviations in the free surface height
are directly balanced by transient changes in the barotropic velocity.
The baroclinic mode meanwhile shows the smoothing of the reduced gravity
pressure, as represented in (\ref{eq:smoothedsw}).

\subsection{The rigid lid assumption}

As we have shown the free surface MGN equations exhibit a strong barotropic
mode, which acts on a much faster scale than the baroclinic modes,
which are the primary interest for studies of internal waves. This
has important implications for numerical calculation, since for time
stepping methods which are conditionally stable it will be the fast
barotropic mode which sets the value of the constraint. For a kilometer
deep ocean basin at hundred meter resolution $\Delta x/\Delta t\gg c=\sqrt{gH}$
requires time steps below one second, much too small for most practical
purposes. One method frequently used to modify the dynamics and remove
this difficulty is to impose a rigid top boundary at $z=0$. Under
the EP framework this can be easily enforced using an additional constraint
term in the reduced Lagrangian,

\[
\int\phi\left[\sum_{i=1}^{N}D_{j}-b\right]d\mathbf{x},\]
with $\phi[x,y]$ a Lagrange multiplier be determined by the constraint
that the term in brackets vanish everywhere . Under application of
the EP methodology an additional pressure term equal to the multiplier
$\phi$, thus constant in all layers, appears in the term due to the
thickness densities, $\delta\ell/\delta D_{i}$. The value of $\phi$
can be obtained through applying the rigid lid condition to the derived
equations and solving the resultant elliptic system. The net effect
of these additional terms is to modify the dispersion relation for
the system, reducing its dimension so that the original fast mode
is removed. The modified dispersion relation is found to be\[
\fl\left[1+\frac{1}{d_{1}+d_{2}}\left(d_{1}^{2}d_{2}\left[\frac{\rho_{1}}{\rho_{2}}+\frac{1}{3}\right]-\frac{d_{1}d_{2}^{2}}{6}-\frac{\rho_{1}d_{1}^{3}}{2\rho_{2}}\right)k^{2}\right]\omega^{2}-g'\frac{d_{1}d_{2}}{d_{1}+d_{2}}=0,\]
which possesses only the baroclinic roots, as stated. This means that
the timestep for numerical methods is then controlled by the large
baroclinic mode representing most of the energy of internal solitary
waves.

We observe that under the rigid lid assumption the no-normal flow
condition requires that the vertical velocity vanishes on the top
surface, so that

\[
w(z=0):=D_{1}\nabla\cdot\mathbf{u}_{1}+W_{1}=0.\]
If this substitution is made in the reduced Lagrangian before the
variations are taken then the resulting equations in the two layer,
one dimensional, case agree exactly with the equations from the multiplier
method, since in one dimension conservation of the MGN PV is automatic.
Both methods produce a system identical to the CC equation for flow
under a rigid lid with varying topography.

\section{Travelling wave solutions of the MGN equations}

The one-dimensional GN equations are widely known to posses a travelling
wave solution \cite{GN74} of the form

\[
D=d\left(1+\left(\frac{c^{2}}{gd}-1\right){\rm sech}^{2}\left(\frac{\sqrt{3(c^{2}-gd)}}{2cd}(x-ct)\right)\right),\]

\begin{equation}
u=c\left(1-\frac{d}{D}\right),\label{eq:trv:u}\end{equation}
where the quantity $c$ is the group speed (and, since these are shallow
water, waves phase speed) of a chosen wave. These waves exist for
all $c$ such that a condition for supercritical flow,

\begin{equation}
\frac{c^{2}}{gd}>1,\label{eq:mincgn}\end{equation}
is satisfied. This also defines a Froude number, $c/\sqrt{gd}$, for
the system. The ${\rm sech}^{2}$ form is also found for the KdV equation,
although the precise definition of velocity differs. The equation
giving the GN travelling wave velocity, (\ref{eq:trv:u}), follows
directly from the transport equation for layer thickness (\ref{eq:trnsp:D}).
Examining the Lagrangian (now calculated against a flat background
state) in the case $N=1$,\[
\ell=\int dxdy\frac{\rho}{2}\left[D|\mathbf{u}|^{2}+\frac{D^{3}}{6}\left(\frac{\partial u}{\partial x}\right)^{2}-g(D-d)^{2}\right],\]
we see the travelling wave solution is a solution of the Hamiltonian
formulation obtained by a direct substitution of the definition of
$u$ for a travelling wave, (\ref{eq:trv:u}). The resulting Lagrangian
is

\[
\mathcal{L}(D,dD/dX)=\frac{\rho}{2}c^{2}D\left(1-\frac{d}{D}\right)-g(D-d)^{2}+\frac{c^{2}}{6D}\left(\frac{dD}{dx}\right)^{2},\]
which under the standard Legende transform, $p=\frac{\partial\mathcal{L}}{\partial\dot{q}}$,
gives a canonical Hamiltonian in terms of $q=D$ of \[
\mathcal{H}(p,q)=\frac{\rho}{2}\left[\frac{c^{2}d^{2}}{6q}p^{2}-c^{2}q\left(1-\frac{d}{q}\right)+g(q-d)^{2}\right].\]
The same equation can also be derived indirectly \cite{Li2002} by
noting that the one-dimensional GN equation possesses a conserved
layer momentum, $\int m/Ddx$, and that travelling waves are stationary
functions of the quantity \[
Q=\int\mathcal{H}-c\frac{m}{D}dx.\]
That both methods give the same final result follows from the invariance
of the original Lagrangian $\ell$ to Galilean boosts, $(x,t)\rightarrow(x-ct,t)$,
so that changing to a frame moving at the constant wave speed requires
only the redefinition of the rest kinetic energy.

\begin{figure}
\begin{centering}
(a)\includegraphics[width=0.8\textwidth]{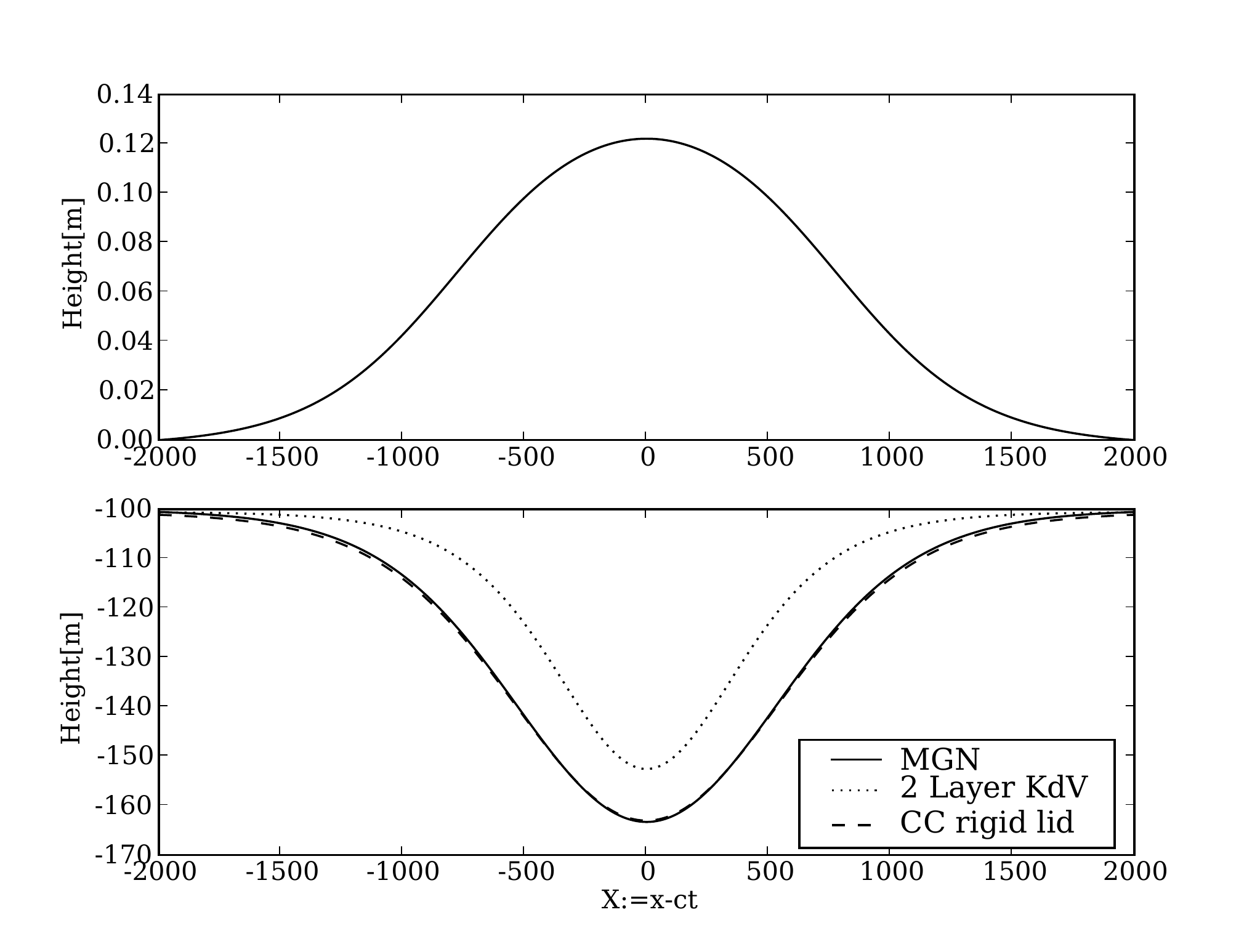}
\par\end{centering}

\begin{centering}
(b)\includegraphics[width=0.8\textwidth]{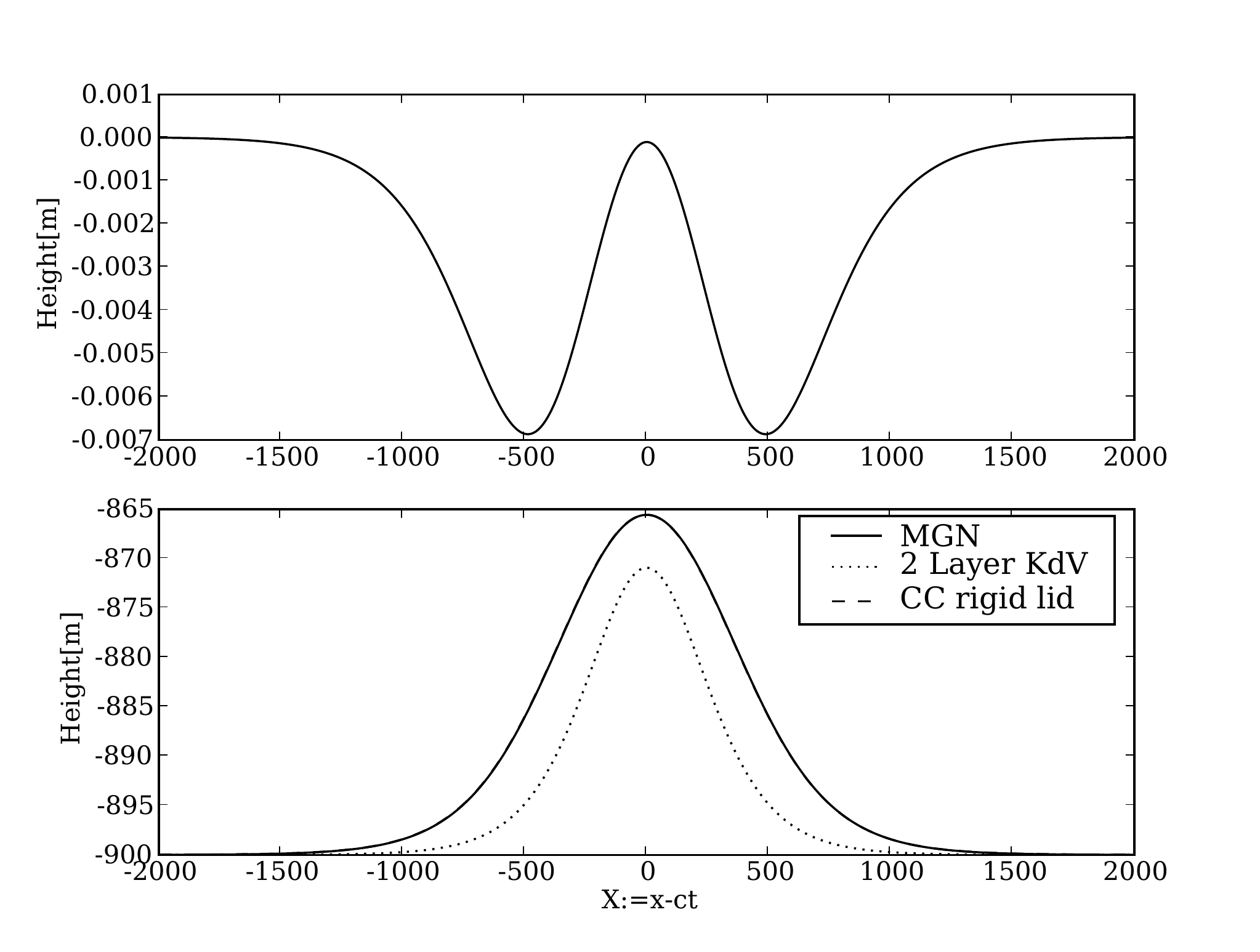}
\par\end{centering}

\caption{Numerical shooting solutions for the MGN travelling wave problem showing
(a) a two layer wave of depression (b) a two layer wave of elevation.
The CC and two layer KdV solutions for the fluid interface are shown
for reference. In both cases the interface of the MGN wave virtually
overlies the CC wave for the chosen wave-speed with the KdV wave noticeably
smaller in magnitude. However the MGN equations also give a free surface
deviation. The apparent dipole structure of the free surface in case
(b) is not evident in the individual layer thicknesses which remain
precisely out of phase.}

\label{fig:TrvSlns}
\end{figure}

The Hamiltonian structure extends easily to the multilayer case, where
each layer possesses an equation identical to (\ref{eq:trv:u}). For
general $N$ the MGN travelling wave solution is a homoclinic orbit
of the Hamiltonian system given by

\begin{eqnarray*}
\fl\mathcal{H}(\mathbf{p},\mathbf{q})=\frac{1}{2}\mathbf{p^{T}\mathbf{A}^{-1}\mathbf{p}}\\
+\frac{1}{2}\sum_{i=1}^{N}\rho_{i}\left[g\left(\sum_{j=i}^{N}(q_{j}-d_{j})^{2}-\sum_{j=i+1}^{N}(q_{j}-d_{j})^{2}\right)-c^{2}q_{j}\left(1-\frac{d_{j}}{q_{j}}\right)^{2}\right],\end{eqnarray*}
around the equilibrium point $(\mathbf{p},\mathbf{q})=(\mathbf{0},\mathbf{d})$,
where the $\mathbf{p}$, $\mathbf{q}$ and $\mathbf{d}$ are the $N$-dimensional
vectors containing the $p_{i}$, $q_{i}$ and $d_{i}$ and $\mathbf{A}$
is an $N\times N$ matrix with elements defined by

\[
A_{ij}=\left\{ \begin{array}{cc}
c^{2}\left(\frac{\rho_{i}d_{i}^{2}}{3D_{i}}+\sum_{k=1}^{i-1}\frac{\rho_{k}d_{k}^{2}}{D_{k}}\right) & i=j\\
\sum_{k=1}^{\min\{i,j\}}c^{2}\frac{\rho_{k}d_{k}^{2}}{2D_{k}} & i\ne j\end{array}\right..\]
Since $\mathbf{A}$ depends on $\mathbf{q}$ the system is not separable
and the dynamics can be extremely complex. Figure \ref{fig:TrvSlns}
shows numerical solutions for the travelling wave problem, calculated
by shooting along the unstable manifold from the equilibrium $(\mathbf{p},\mathbf{q})=(\mathbf{0},\mathbf{d})$
at $x=-\infty$. The integration problem was found to be extremely
stiff and the use of a symplectic integration method, the generalized
leapfrog method, was implemented to maintain stability. For comparison,
we also plot the relevant numerical solutions to the CC equations
\cite{CC96JFM} and the algebraic solution to the two-layer KdV equation
\cite{Crighton1995}. 

\begin{figure}
\begin{centering}
\begin{tabular}{cc}
(a) & \includegraphics[width=0.75\textwidth]{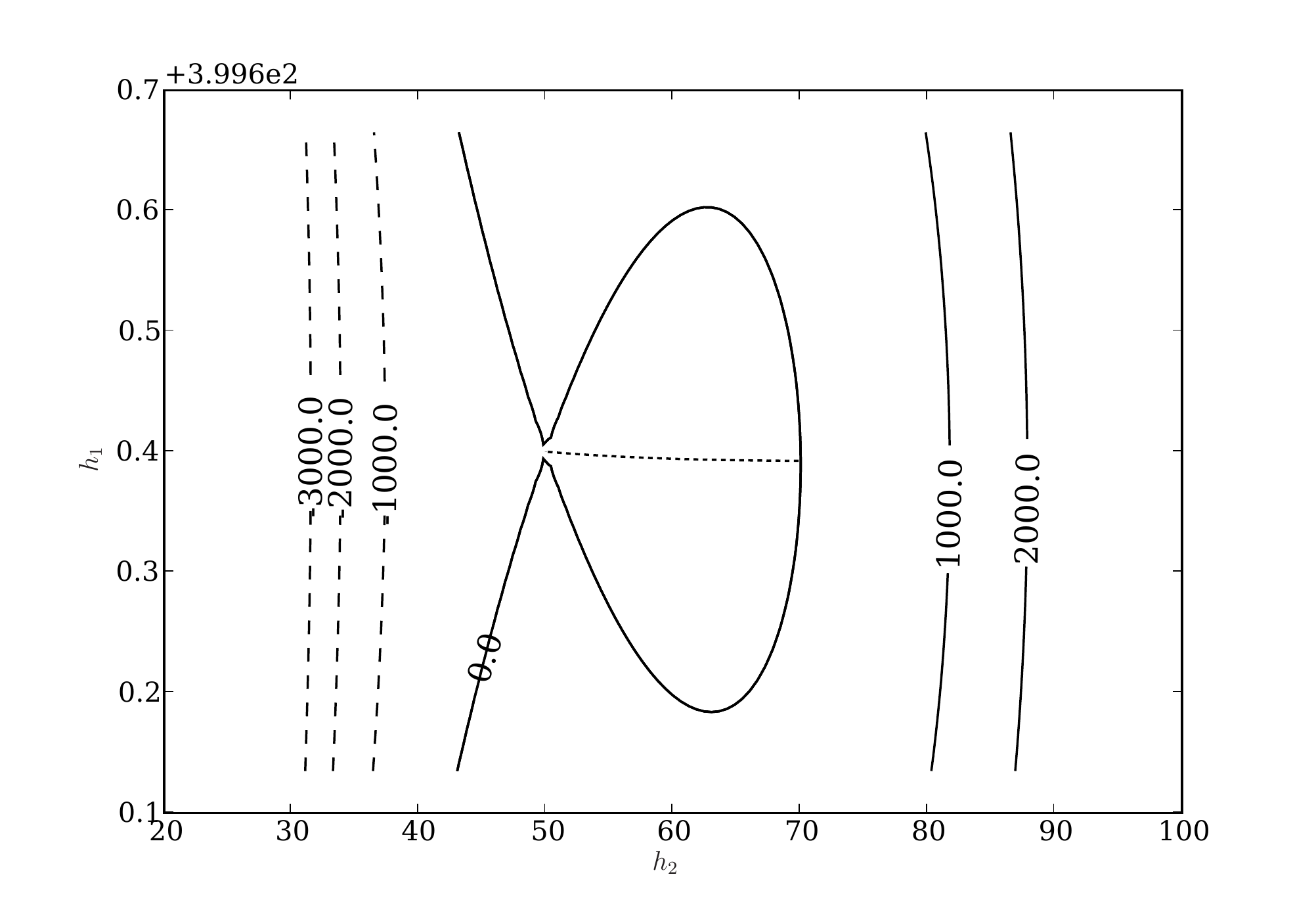}\tabularnewline
(b) & \includegraphics[width=0.75\textwidth]{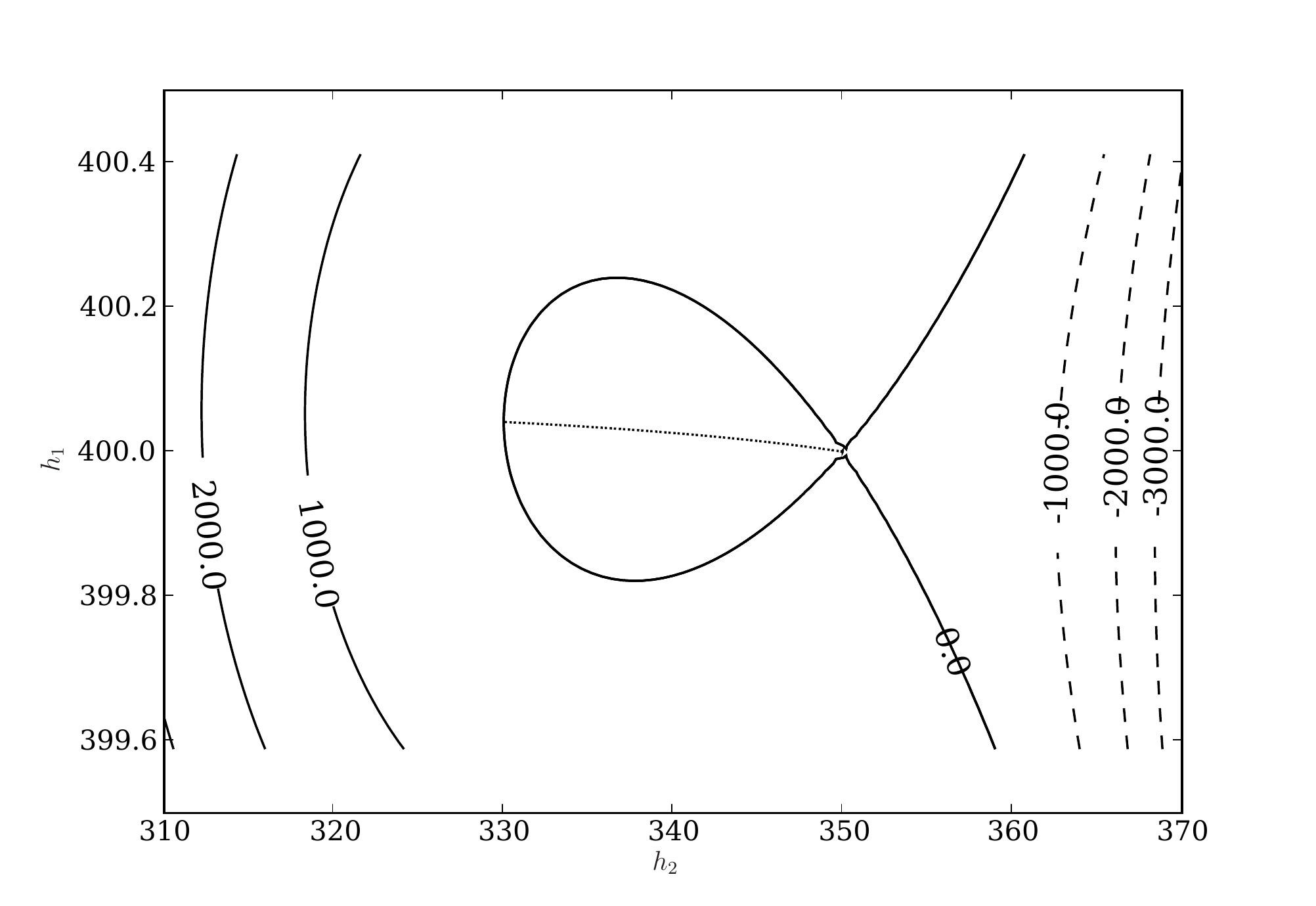}\tabularnewline
(c) & \includegraphics[width=0.75\textwidth]{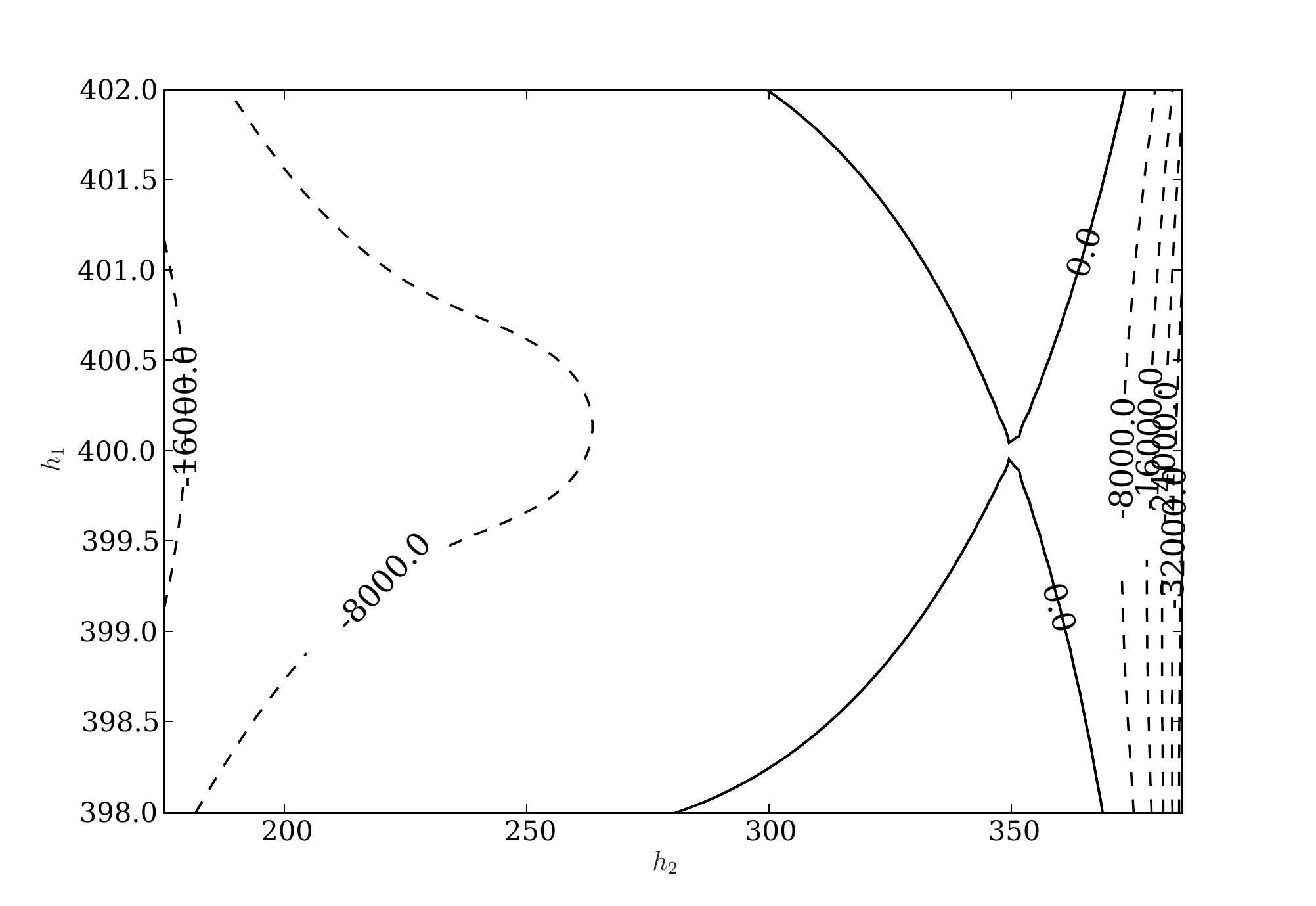}\tabularnewline
\end{tabular}
\par\end{centering}

\caption{Contour plots of potentials which allow (a) a wave of elevation, (b)
a wave of depression (c) no travelling wave solutions. In the first
two cases the trajectory for the travelling wave solution of the particular
regime is plotted as a dotted line, contained within the zero contour.
The fish-like looped zero contours disappear with increasing wave
speed through a bifurcation with a second $(h_{1},h_{2})$ contour
joining $(\infty,\pm\infty)$ to $(-\infty,\pm\infty)$ with the sign
of $h_{2}$ positive for waves of elevation and negative for waves
of depression.}
\label{fig:pots}
\end{figure}

The condition for supercritical flow and thus existence (\ref{eq:mincgn})
generalises to a condition that the stable manifold of the system
at $x=-\infty$ must be of dimension greater than zero. This is equivalent
to the condition that $\det(\mathbf{V}-\lambda^{2}\mathbf{A})=0$
has real solutions, where $\mathbf{V}$ is the matrix \[
V_{ij}=-\left.\frac{\partial^{2}\mathcal{H}}{\partial q_{i}\partial q_{j}}\right|_{(\mathbf{p},\mathbf{q})=(\mathbf{0},\mathbf{d})}=\left\{ \begin{array}{cc}
\rho_{i}\left(\frac{c^{2}}{d_{i}}-g\right) & i=j\\
-g\rho_{\min\{i,j\}} & i\ne j\end{array}\right..\]
This matrix comes from the form of the Jacobian matrix of the Hamiltonian
system linearized around the far field values. For $N=2$ the critical
condition for internal and external travelling waves become respectively

\begin{equation}
\frac{c^{2}}{g}>\frac{d_{1}+d_{2}\mp\left((d_{1}+d_{2})^{2}-4(\rho_{2}-\rho_{1})d_{1}d_{2}/\rho_{2}\right)^{1/2}}{2}.\label{eq:minc}\end{equation}
This is however only a necessary condition on the wave speed and there
exist stratifications for which no internal travelling wave is possible
for $c$ supercritical. This can be illustrated by considering the
form of the potential\[
\fl\mathcal{V}(\mathbf{q})=\frac{1}{2}\sum_{i=1}^{N}\rho_{i}\left[g\left(\sum_{j=i}^{N}(q_{j}-d_{j})^{2}-\sum_{j=i+1}^{N}(q_{j}-d_{j})^{2}\right)-c^{2}q_{j}\left(1-\frac{d_{j}}{q_{j}}\right)\right]\]
 in the Hamiltonian in the regimes where there exist waves of elevation,
waves of depression, and for cases with no travelling wave. Contour
plots of the potential in these three conditions are shown in Figure
\ref{fig:pots}, along with the marked trajectories where a travelling
wave exists. The potential is seen to undergo bifurcations with the
creation or destruction of homoclinic contours around the equilibrium
at $\mathbf{q}=\mathbf{d}$. The direction of the contour defines
whether the wave is of elevation or depression and its destruction
with increasing wave-speed represents a limit on the speed (and thus
amplitude) of travelling waves allowed by the system. This follows
since for N=2 the matrix $\mathbf{A}$ is positive definite and thus
$\mathbf{p}^{T}\mathbf{A}^{-1}\mathbf{p}$ is a strictly positive
quantity. Jo and Choi \cite{JoChoi2002} investigate the two-layer
system in the rigid lid case and find conditions on minimum and maximum
wave speed similar to those presented here. The critical flow condition
is

\[
\frac{c^{2}}{g}>\frac{(\rho_{2}-\rho_{1})d_{1}d_{2}}{\rho_{2}(d_{1}+d_{2})},\]
this is the first order term in the Taylor series expansion of (\ref{eq:minc})
in the limit $(d_{1}+d_{2})^{2}\gg4(\rho_{2}-\rho_{1})d_{1}d_{2}/\rho_{2}$.
This suggests that the rigid lid model is a good approximation when
density differences are small, or the aspect ratio $d_{1}/d_{2}$
differs greatly from unity, as may be expected. A similar set of calculations
and analysis is in progress for the case $N=3$, which appears to
increase the dimension of possible behaviour. 

\section{Summary}

We have introduced a set of equations derived from a variational principle
under an ansatz of columnar motion. These have been shown to be identical
to the multilayer Green Nagdhi equations derived independently by
other researchers by other methods. These equations are proved to
contain a fast barotropic mode which is virtually unmodified by the
nonlinear part of the MGN operator. This means the equations require
careful handling in numerical simulations. The travelling wave solutions
are shown to also be derivable from a variational principle, and to
show a more complex range of behaviour and waveforms than the single
layer or rigid lid cases.

\ack{}{}

We thank W. Choi and R. Grimshaw for fruitful discussions of this
work and their own and two anonymous reviewers for their insightful
comments. J. R. Percival was supported by the US ONR under grant N00014-05-1-0703.

\bibliographystyle{unsrt}
\bibliography{MLCM}

\end{document}